\documentclass[prl, twocolumn]{revtex4-1}
\usepackage{amsmath}
\usepackage{amssymb}
\usepackage{graphicx}
\usepackage{color}
\usepackage{multirow}
\newcommand{\be}{\begin{equation}}
\newcommand{\ee}{\end{equation}}
\newcommand{\bea}{\begin{eqnarray}}
\newcommand{\eea}{\end{eqnarray}}
\newcommand{\balg}{\begin{align}}
\newcommand{\ealg}{\end{align}}

\begin{document}
  \title{Analysis of Over-magnetization of Elemental Transition Metal Solids 
from the SCAN Density Functional}
  \author{Daniel Mej{\'i}a-Rodr{\'i}guez}
  \email{dmejiarodriguez@ufl.edu}
  \author{S.B.\ Trickey}
  \email{trickey@qtp.ufl.edu}
  \affiliation{Quantum Theory Project, Department of Physics, University of Florida, Gainesville, FL 32611}
  \date{21 June 2019}
	
\begin{abstract}
\noindent Recent investigations have found that the strongly constrained and
appropriately normed (SCAN) meta-GGA exchange-correlation functional
significantly over-magnetizes elemental Fe, Co, and Ni solids.  For
the paradigmatic case, bcc Fe, the error relative to experiment is
$\gtrsim 20 \%$.  Comparative analysis of magnetization results from
SCAN and its \textit{deorbitalized} counterpart, SCAN-L,
leads to identification of the source of the discrepancy. It is not
from the difference between Kohn-Sham (SCAN-L) and generalized Kohn-Sham (SCAN)
procedures. The key is the iso-orbital 
indicator $\alpha$ (the ratio of the local
Pauli and Thomas-Fermi kinetic energy densities).  Its 
\textit{deorbitalized} counterpart, $\alpha_L$, has more dispersion
in both spin channels with respect to magnetization 
in an approximate region between 0.6 Bohr and 1.2 Bohr
around an Fe nucleus. The overall effect is that
the SCAN switching function evaluated with $\alpha_L$ reduces
the energetic disadvantage of the down channel with respect to up compared to
the original $\alpha$, which in turn reduces the  magnetization.  This
identifies the cause of the SCAN magnetization error as insensitivity 
of the SCAN switching function to $\alpha$ values in the approximate range
$0.5 \lesssim \alpha \lesssim 0.8$ and oversensitivity for $\alpha \gtrsim 0.8$.
\end{abstract}

  \maketitle
Transition metals generally - and Fe 
particularly - are central to both practical applications and 
to the development of improved exchange-correlation (XC) approximations
for use in density functional calculations.  
A pertinent example of the latter role is the generalized 
gradient approximation (GGA) \cite{PW91a,PW91b}
breakthrough that gave the right ground-state crystal structure and
magnetic order for elemental Fe \cite{SinghPickettKrakauer1991}. 
In light of that importance, some results from a very sophisticated 
meta-GGA functional are quite provocative.

Several authors have found that the Strongly Constrained and
Appropriately Normed (SCAN) \cite{SCAN,SCANNature} meta-GGA XC functional 
over-magnetizes some elemental transition metals \cite{IsaacsWolverton2018,%
JanaPatraSamal18,RomeroVerstraete2018,EkholmEtAl2018,FuSingh2018,FuSingh2019}.
For example, Isaacs and Wolverton \cite{IsaacsWolverton2018} found
that SCAN over-magnetizes bcc Fe by 19 \%, hcp Co by  8\% and
fcc Ni by 14\%. They also found that ``\ldots [based on local moment 
calculations] the average maximum magnetic 
moment within SCAN is 12\% larger than that found within PBE.''
Data from Jana et al.\ \cite{JanaPatraSamal18} 
correspond to magnetization excesses of 20 \% (bcc Fe), 4 \% (fcc Co), 
and 8 \% (fcc Ni).
In the ordered 50-50 Fe$_3$Pt alloy, Romero and Verstraete 
\cite{RomeroVerstraete2018} found that
SCAN gave  $\approx 12$\% over-magnetization on the Fe site compared 
to 3\% overage from PBE.
Ekholm et al.\ \cite{EkholmEtAl2018} found excesses of 25\% for bcc Fe, 
9.5\% for hcp Co, and 28\% for fcc Ni.

Most recently, Fu and Singh \cite{FuSingh2018} showed that 
SCAN over-magnetizes bcc Fe by $\approx 23\%$, hcp Co by
14\% and fcc Ni by 13\%. In Fe$-3$C, they found SCAN gives 
nearly 30\% over-magnetization per three-iron-atom formula unit.
Subsequently they \cite{FuSingh2019} have 
suggested that the SCAN functional is intermediate between PBE and approaches
that describe more localized systems better 
(e.g., hybrid functionals or DFT+$U$), hence SCAN 
tends to yield over-magnetization.

Overall, the trend is completely clear.  SCAN over-magnetizes
elemental transition metal solids.
Presumably the numerical differences 
(compared for the same crystalline phases)
trace to differences in computational parameters and techniques and to
the intrinsic sensitivity of magnetization calculations. 
 Given the other broad successes of 
SCAN (e.g.\ Ref. \cite{IsaacsWolverton2018} and references therein),
the discrepancy is noteworthy.  SCAN results for these simple systems
are strikingly different from the behavior found from other meta-GGA
functionals (e.g.\ TPSS \cite{TPSS}) or a typical GGA (e.g.\ PBE
\cite{PBE}). Both give close to the experimental magnetization.

The unresolved issue is the specific source of
the discrepancy: what within the SCAN functional leads to such 
strikingly different magnetization behavior compared to other
semi-local functionals? With the aid of SCAN-L \cite{SCANL1,SCANL2},
our \textit{deorbitalized} version of SCAN, we can resolve the issue
and, as well, provide insight perhaps useful for the development of
better meta-GGA functionals.

A few definitions are useful.  SCAN uses the so-called iso-orbital indicator
\be
\alpha(\mathbf{r}):= (\tau_s -\tau_W)/\tau_{TF} \; .
\label{alphadefn}
\ee
Here $\tau_s = (1/2)\sum f_j \vert \nabla \varphi_j(\textbf{r})\vert^2$ is
the positive-definite Kohn-Sham (KS) kinetic energy density in terms
of the KS orbitals $\varphi_j$ (with occupations $f_j$), and $\tau_W$ and $\tau_{TF}$ are the
von Weizs\"acker and Thomas-Fermi kinetic energy densities
respectively. The numerator of $\alpha$ is the Pauli KE density.  It
vanishes for the case of a single-orbital system, one of the ways that
$\alpha$ enables a functional to distinguish chemically different
bonding regions.  The deorbitalized SCAN, SCAN-L, differs only in
using an approximate orbital-independent $\alpha [n, \nabla n, \nabla^2n]$ 
with $n$ the electron number density.

Turning to analysis, first we can eliminate the possibility that the
SCAN magnetization discrepancies 
arise from limitations of computational technique. The potential 
issue is that PAW data sets do not exist for SCAN (nor for other
meta-GGAs).  Thus the VASP calculations \cite{vasp3,vasppaw} reported here 
and earlier 
\cite{IsaacsWolverton2018,EkholmEtAl2018,FuSingh2018,FuSingh2019} used PBE PAWs
instead.  However, in addition to those calculations,
two groups \cite{EkholmEtAl2018,FuSingh2018,FuSingh2019} also did post-scf
all-electron calculations with the WIEN2k code \cite{WIEN2k} and PBE
spin densities and found the same distinctive over-magnetization trend 
for SCAN.

Therefore the issue originates structurally in SCAN. An obvious 
question is whether SCAN-L, which uses the same structure as SCAN,
inherits the over-magnetization. Table \ref{CalcVals} 
  \begin{table}
\caption {Calculated bcc Fe lattice parameters, saturation magnetizations, 
and FSM energies for various XC functionals at $a_{calc}$. \label{CalcVals}}

    \begin{tabular}{l c c c}
              & $a_{calc}$ (\AA) & $m_{sp}$ ($\mu_B$/atom) & $E_{mag}$ (meV/atom) \\\toprule
     PBE      &  2.82       &    2.14                 & -564                 \\             
     TPSS     &  2.80       &    2.12                 & -645                 \\             
     SCAN     &  2.85       &    2.60                 & -1100                \\             
     regSCAN  &  2.84       &    2.62                 & -1201                \\             
     SCAN-L   &  2.81       &    2.05                 & -653                 \\             
     TPSS-L   &  2.81       &    2.09                 & -568                 \\\toprule     
    \end{tabular}
\vspace*{-12pt}
  \end{table}
shows results of VASP \cite{vasp3}
calculations (with PAWs \cite{vasppaw} and other parameters as in Ref.\ 
\cite{SCANL2}). SCAN and SCAN-L yield qualitatively different
magnetizations.  Consistent with prior results 
summarized above, in our fixed spin moment (FSM) calculations   
SCAN gives an overly stable bcc Fe structure
that is over-magnetized: $m_{sp} =2.60 \mu_B$ at 
FSM energy  $\vert E_{mag}\vert = 1100$ meV below 
zero moment at calculated equilibrium lattice parameter.
For comparison, the  PBE values are $2.21 \mu_B$ and 564 meV. 
\begin{figure} [h]
	\includegraphics[width=0.9\linewidth]{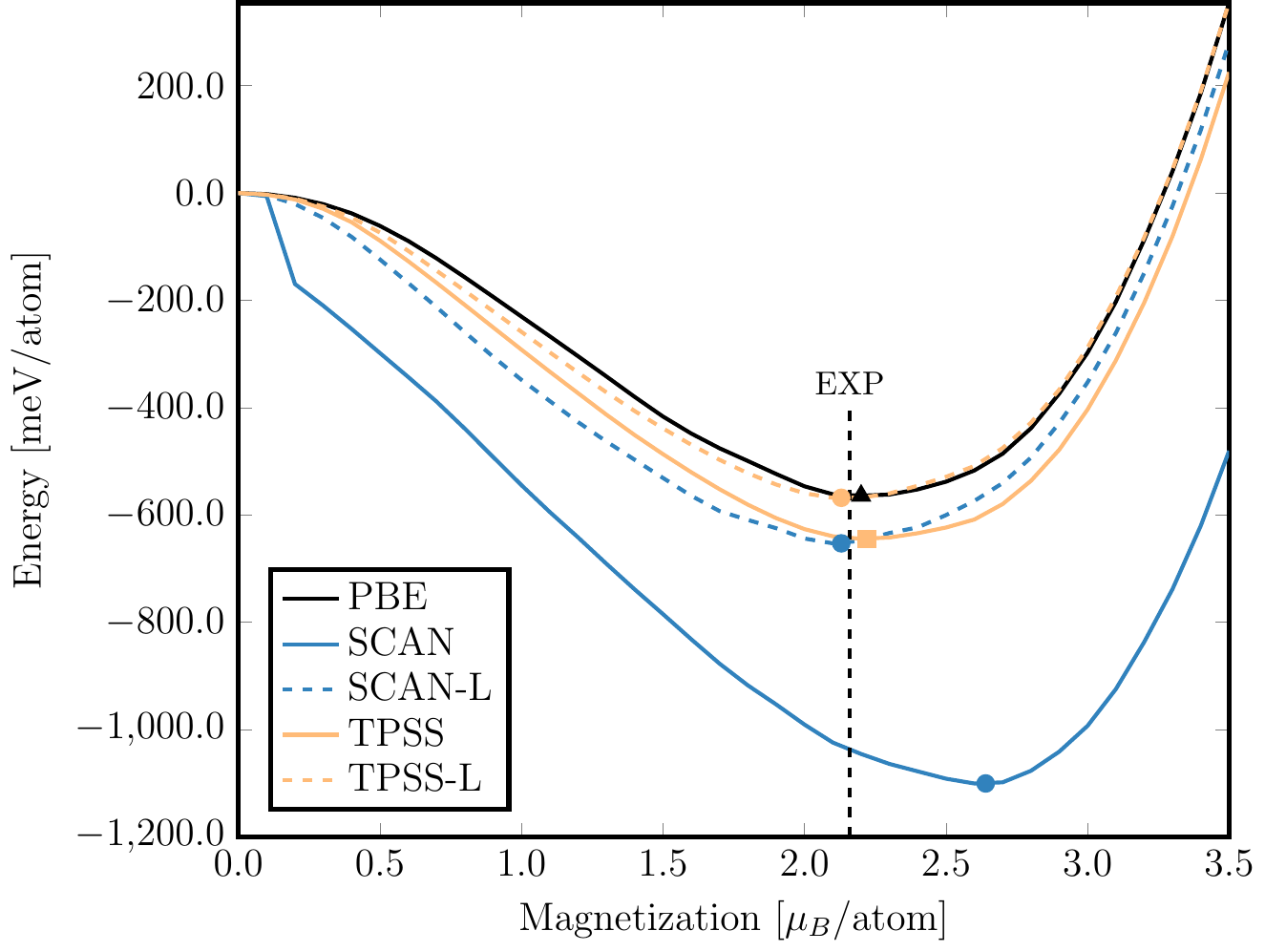}
	\caption{Fixed spin-moment energy versus magnetic moment 
for bcc Fe at $a_{exp}$ from PBE, SCAN, SCAN-L, TPSS and TPSS-L.  
Dots show minimum FSM energy values.  Experimental
value from Ref.\ \onlinecite{FuSingh2018}.\label{fig:FeMag}}
\vspace*{-6pt}
\end{figure}

Fig.\ \ref{fig:FeMag} (the counterpart to 
Fig.\ 1 of Ref.\ \onlinecite{FuSingh2018}) shows the dramatic difference in FSM energy as a function 
of magnetization for SCAN compared to other functionals. 
In contrast to the SCAN magnetization,
SCAN-L reduces both $\vert E_{mag}\vert$ and $m_{sp}$ to the
point of being essentially the same as the PBE results. 
(Aside: the kinked behavior of the SCAN $\vert E_{mag}\vert$ 
at  $m \approx 0.3 \mu_B$ seems to arise from numerical instabilities.
It does not appear in the post-scf SCAN curve in Fig.\ 1 of 
Ref.\ \onlinecite{FuSingh2018} and is immaterial to the issue at hand.)  

Upon first thought, the origin of the difference between SCAN and
SCAN-L magnetization might be suspected to have arisen from the procedural
difference associated with use of orbital-dependent and
orbital-independent functionals.  The orbital independence of SCAN-L
leads to a multiplicative XC potential and use of the ordinary
Kohn-Sham (KS) procedure.  The orbital-dependent SCAN XC potential, in
contrast, almost always is used in the generalized Kohn-Sham (gKS)
context. The two schemes are inequivalent \cite{YangPengSunPerdew}.
However, $m_{sp}(a_{calc})$ values for bcc Fe from the TPSS functional
\cite{TPSS}, obtained with gKS, and from its deorbitalized version,
TPSS-L, which uses KS, do not exhibit the remarkable difference of the
SCAN vs.\ SCAN-L case.  TPSS and TPSS-L deliver $m_{sp}$ values
indistinguishable from each other and from the PBE result (see
Figure\ \ref{fig:FeMag}).

The post-scf calculations quoted already 
\cite{EkholmEtAl2018,FuSingh2018,FuSingh2019} in fact confirm 
the irrelevance of KS and gKS for over-magnetization.
The sole difference between the post-scf SCAN energy and the underlying 
PBE energy is the exchange-correlation energy
$E_{xc}$ evaluation. The two calculations differ only by
$E_{xc}^{SCAN}[\lbrace \phi_j^{PBE}\rbrace] - E_{xc}^{PBE}[n_{PBE}]$. They 
are evaluated entirely with KS quantities.  There are no
gKS inputs.  
Nevertheless, in the post-scf calculations SCAN
overmagnetizes.  In this context, it also is worth mentioning that
those post-scf results are incompatible with the localization
explanation offered in Ref.\ \onlinecite{FuSingh2019}, since PBE
calculations suffer from delocalization error due to 
self-interaction. (The extent, if any, to which SCAN causes localization 
to occur in the self-consistent VASP 
calculations reported here and earlier is indeterminate at this point.  
That follows from the inherent spurious delocalization in the PBE PAWs
used. The issue is irrelevant to the analysis below.)

\begin{figure} [h]
	\includegraphics[width=0.8\linewidth,trim={1.2cm 0 1.2cm 0}]{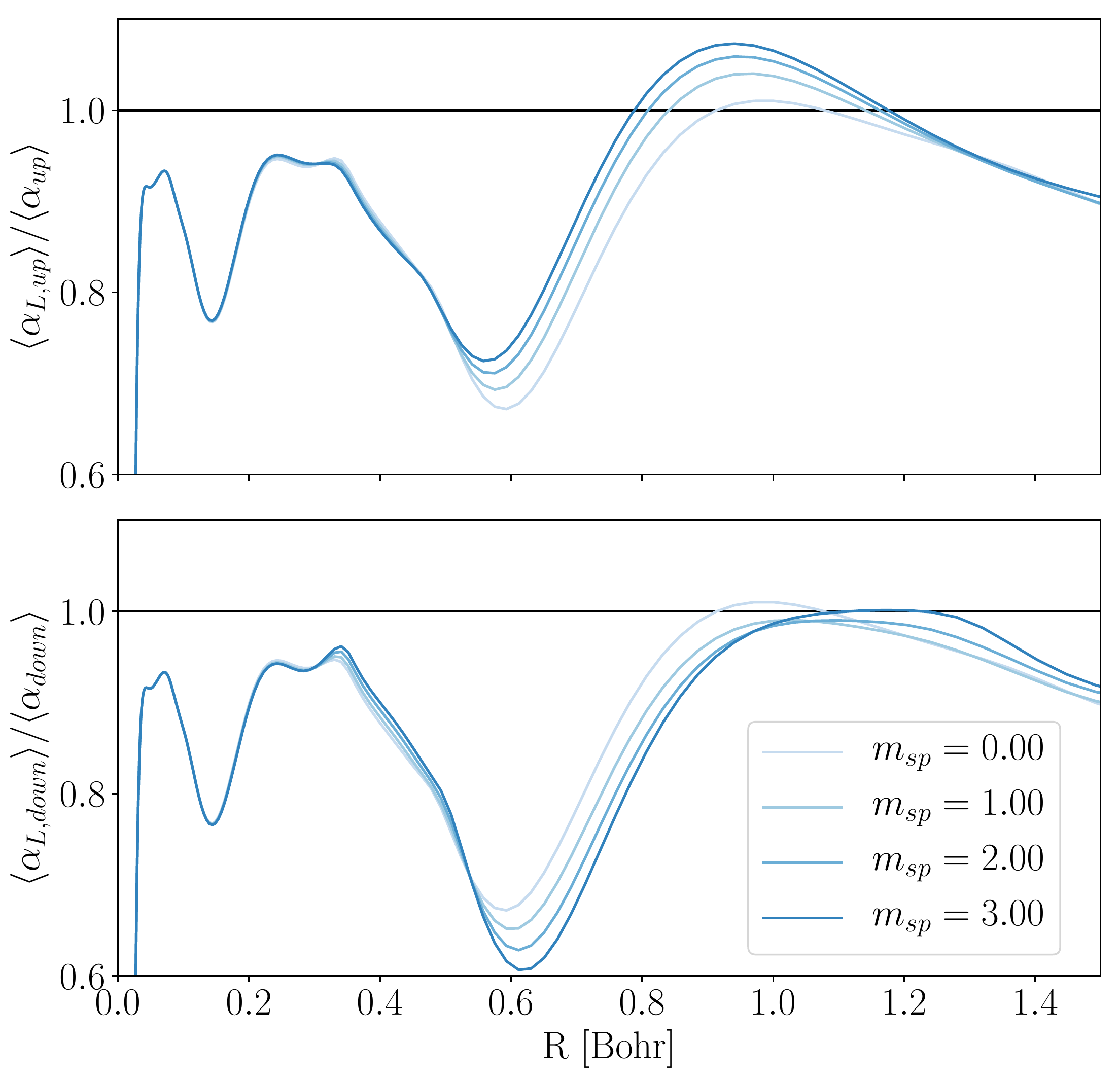}
	\caption{Ratio of angularly averaged $\alpha$s, $\langle \alpha_L\rangle/ \langle \alpha \rangle$ for
spin-up (above) and spin-down (below) as function
of radial distance from Fe nucleus (evaluated with bcc Fe PBE densities) at
selected fixed spin moments.  
\label{fig:AlphaRatio}}
\vspace*{-12pt}
\end{figure}

The only other difference between SCAN and SCAN-L is the distinction
between $\alpha$ and $\alpha_L$. In most
cases, $\alpha_L \approx \alpha$ is very accurate, but there can be
 regions where their difference is noticeable \cite{SCANL1,Tran}.  
That is critical to the present analysis. 
Fig.\ \ref{fig:AlphaRatio} shows the ratio of the angularly averaged
$\alpha_L$ to $\alpha$ for various magnetizations as a function of
distance from the Fe nucleus in bcc Fe (calculated post-scf with bcc
Fe spin densities from PBE).
\begin{figure} [h]
	\includegraphics[width=0.8\linewidth,trim={1.2cm 0 1.2cm 0}]{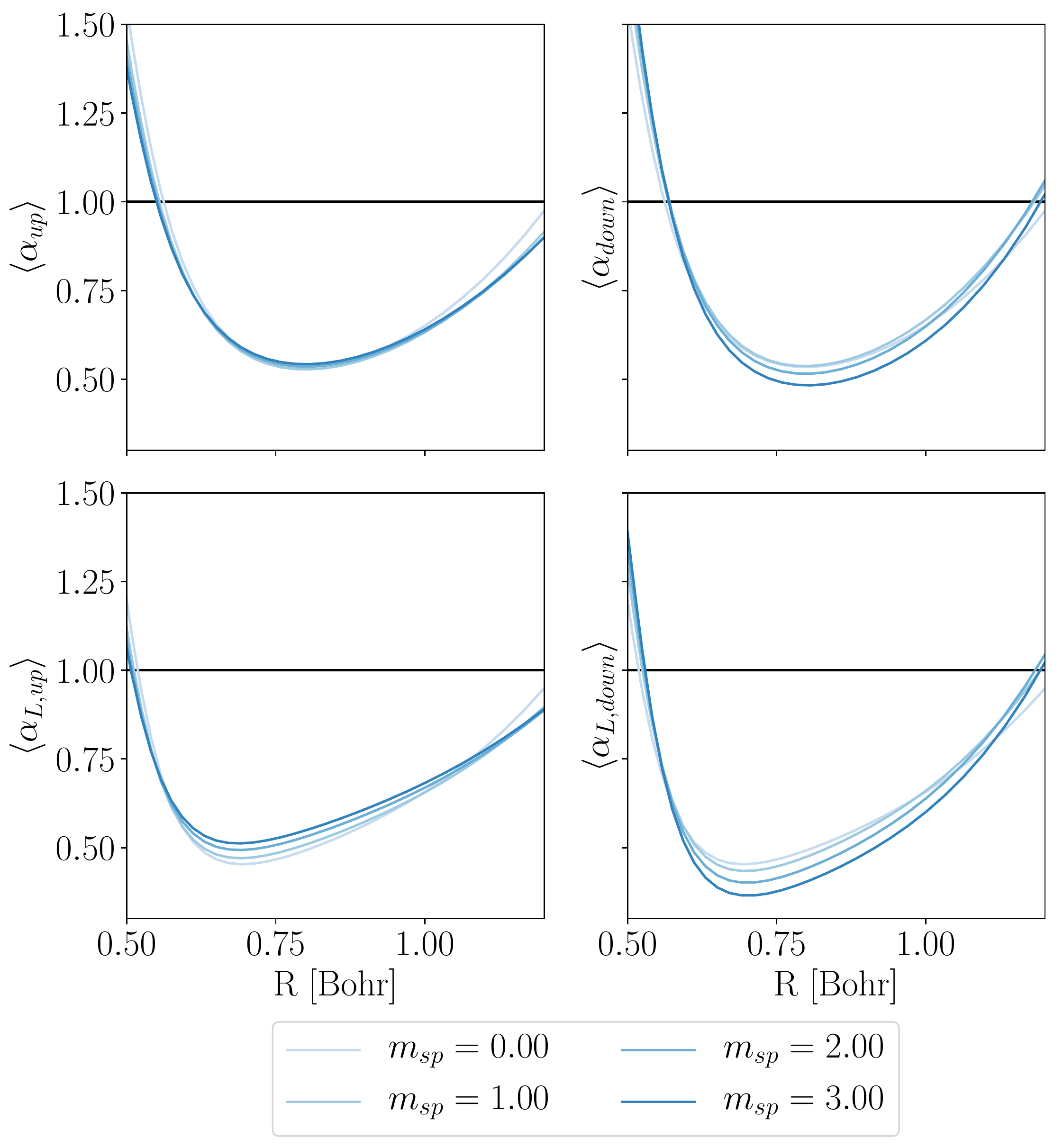}
	\caption{Angularly averaged $\alpha$ (above) and $\alpha_L$ (below)
as a function of radial distance from a Fe nucleus (evaluated with bcc Fe PBE densities) at
selected fixed spin moments. Spin-up plots are in the left column, spin-down in the right.
\label{fig:Alphas}}
\vspace*{-12pt}
\end{figure}
Key distinctions to note include the fact that
below about $r=$0.9 bohr, $\alpha_L$ is smaller than $\alpha$ for both
spins, with particularly strong reduction on $0.6 \lesssim r \lesssim 0.8$ bohr. 
It also is important that the ratios for both spins
exhibit some dispersion with respect to $m_{sp}$, especially in
the same $0.6 \lesssim r \lesssim 0.8$ bohr region.  
Fig.\ \ref{fig:Alphas} shows that such dispersion is present
in both spin channels of $\alpha_L$, but only in
the spin-down channel of $\alpha$. Moreover, $\alpha_{L,down}$ is
more dispersed than  $\alpha_{down}$.

The dominant $\alpha$-dependent contribution to exchange in SCAN and
SCAN-L is from the switching function $f_x(\alpha)$ that distinguishes
regions of $\alpha < 1$ and $\alpha > 1$.
\begin{figure} [h]
	\includegraphics[width=0.9\linewidth,trim={1.2cm 0 1.2cm 0}]{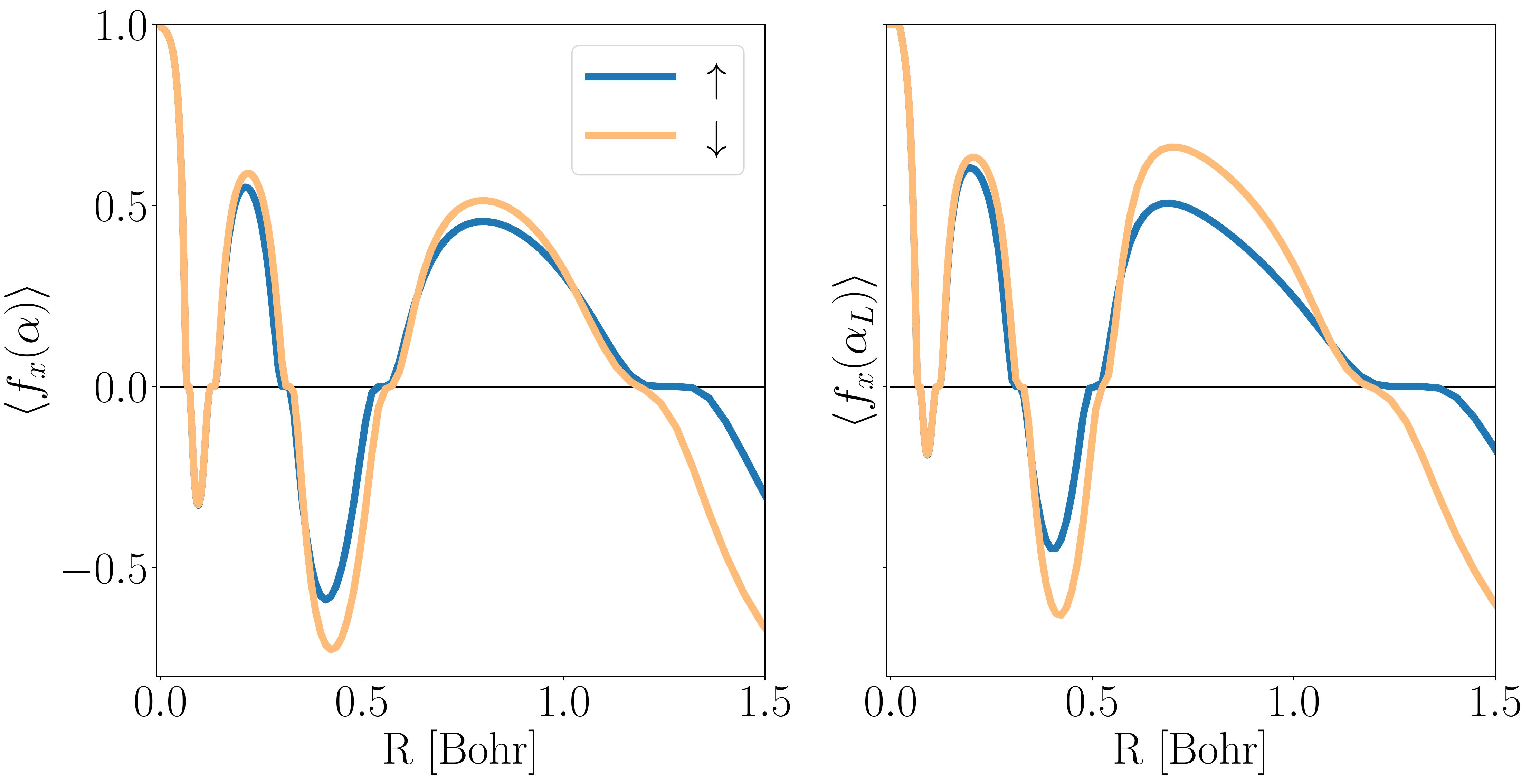}
	\caption{Left: Angularly averaged SCAN switching function, $f_x(\alpha)$, as function
	of radial distance from Fe nucleus (evaluated with bcc Fe PBE spin densities) at
    $m_{sp}=2.5$ $\mu_B/$atom.
	Right: Radial behavior of angularly averaged SCAN switching function evaluated with 
$\alpha_L$, $f_x(\alpha_L)$, for same densities.\label{fig:fswitchFx}}
\vspace*{-6pt}
\end{figure}
An important  bit of analysis is that recently it has
been shown \cite{regSCAN} that modifications to make $f_x$ smoother 
around $\alpha \approx 1$ have negligible effect on SCAN 
structural and energetic predictions 
in solids. Our calculations with that ``regularized SCAN'' (regSCAN) 
confirm similarly little effect of those modifications upon 
the over-magnetization values of $m_{sp}$ and a
9 \% increase on $|E_{mag}|$ (see Table \ref{CalcVals}). 

Fig.\ \ \ref{fig:fswitchFx} shows the angularly averaged
switching function as a function of radial distance evaluated
with $\alpha$ and $\alpha_L$ for both spin channels.
One sees that on $0.6 \lesssim r \lesssim 0.9$ bohr or so, 
the $\alpha_L$ values separate the up and down-spin points on the
$f_x(\alpha)$ curve more than the original $\alpha$ does. In particular, because the 
down-spin ratio $\alpha_L$ to $\alpha$ is below the up-spin ratio in that radial
domain, the down-spin exchange energy density $e_{x,down}$ contributes more
to the full $e_{x}$ for SCAN-L than for SCAN.  The 
dispersion ordering with respect to magnetization of $\alpha_{L,up}$ values
is reversed compared to $\alpha_{L,down}$.  That is, for up spin, greatest
magnetization has the least reduction 
(largest $ \langle \alpha_{L,up} \rangle/\langle \alpha_{up}  \rangle$) 
while for down spin, greatest magnetization has greatest reduction 
(least $ \langle \alpha_{L,down} \rangle/\langle \alpha_{down} \rangle$).  
Added to this is the fact that the up-spin $\alpha$ is 
almost insensitive to $m_{sp}$, hence so is the up-spin $e_x$, which is not the
case for $\alpha_L$.

In the immediately adjacent region, $0.9 \lesssim r \lesssim 1.2$ bohr,
$\alpha_{down} > \alpha_{up}$ and $\alpha_{L,down} > \alpha_{L,up}$. However,
the $\alpha_L$s are closer together. As a result, 
 $e_{x,down}$ also contributes more to the full $e_{x}$ for SCAN-L than for SCAN
in this region.

\begin{figure} [h]
	\includegraphics[width=0.8\linewidth,trim={0 1.0cm 0 1.0cm},clip]{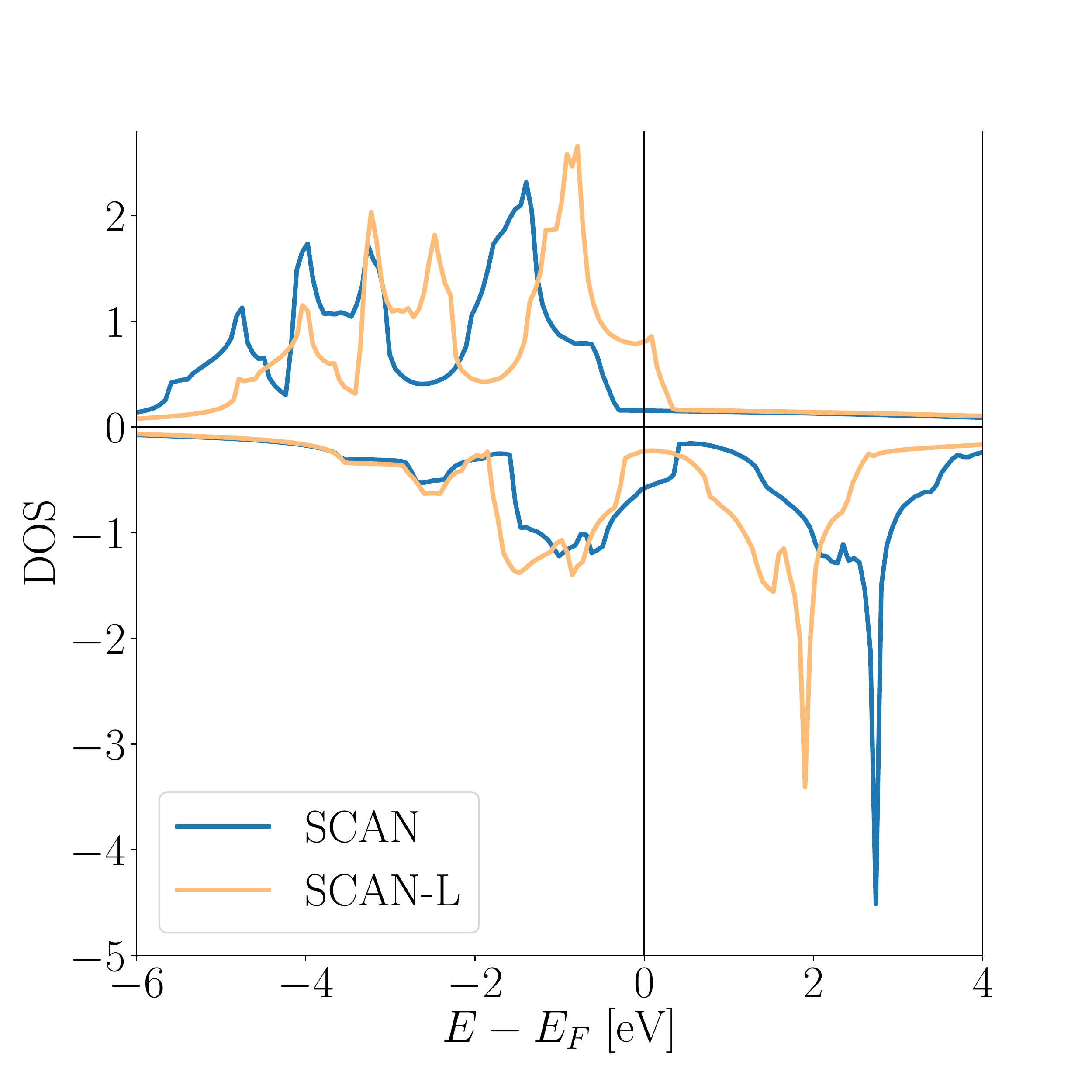}
	\caption{bcc Fe densities of states for SCAN and SCAN-L. Up spin 
upper panel, down spin, lower.\label{fig:BCCFeDOS}}
\vspace*{-2pt}
\end{figure}

The net result is a set of significant 
differences in the densities of states of 
the ground states predicted by SCAN and SCAN-L.  Fig.\ \ref{fig:BCCFeDOS}  
shows that, relative to SCAN, SCAN-L shifts the up-spin occupied states 
up somewhat,  thereby reducing the magnetization energy and leaving their 
population reduced. Meanwhile, the down-spin state energies are
somewhat lowered, corresponding to enhanced $e_{x,down}$ and their population 
therefore goes up. Those shifts may also be related to the increased exchange 
splittings discussed in Ref. \onlinecite{FuSingh2019}.

Since discrepancies of $\alpha_L$ with respect to $\alpha$ identify
the region $0.6 \lesssim r \lesssim 1.2$ bohr as critical, we tried 
a simple modification of the SCAN switching function solely to probe how it 
responds to the actual
orbital-dependent $\alpha$ values in that region.  The deorbitalized 
$\alpha_L$ values for up and down spin are more separated than
the original $\alpha$s in the $0.6 \lesssim r \lesssim 0.9$ bohr region (region A),
but closer together in the $0.9 \lesssim r \lesssim 1.2$ bohr one (region B). 
Another important fact is that $\alpha$ and $\alpha_L$ values 
in both such regions are below 1, but get closer to 1 in region B.
An important probe therefore is to bring the values of the switching 
function for $\alpha_{up}$ closer to the ones for
$\alpha_{down}$ in region B, while separating the switching function values
in region A. This amounts to investigating the consequences of 
reducing the sensitivity of the switching function  to small changes 
in its argument around $0.8 \lesssim \alpha \leq 1$ and increasing
that sensitivity for $\alpha\lesssim$ 0.8. 

For $\alpha < 1$ (see eqs.\ (58)-(60) in
Ref.\ \onlinecite{SCANL2}), the switching function is $f_x(\alpha) =
\exp{ \left[-c_{1x}\alpha/(1-\alpha) \right]}$ with $c_{1x}=0.667$.
A crude way to probe the apparently desirable sensitivity change  
for region B is to increase $c_{1x}$ from 0.667 to 1. 
Doing so yields improved $m_{sp}$ and $E_{mag}$ values for
bcc Fe, but produces mixed results for other systems.
Detailed pursuit of more sophisticated modifications would
be tantamount to constructing a revision or successor to SCAN,
a task far beyond the scope of the present investigation.

What the simple probe and the comparative analysis of $\alpha$ 
and $\alpha_L$ behavior make clear is the significant need 
for more refined switching-functions in improved meta-GGA functionals.
In addition to the over-magnetization origin diagnosed here, 
the SCAN switching-function can be linked to issues of 
numerical integration sensitivity \cite{YangPengSunPerdew}
and self-consistent field stability \cite{regSCAN}.  To that point, it is perhaps suggestive that 
 the Tao and Mo meta-GGA XC functional \cite{TaoMo},
which uses a very different switching function, 
does not give over-magnetization \cite{JanaPatraSamal18}.
 \begin{table}
\caption {Co, Ni and V calculated saturation magnetizations and FSM energies
for various XC functionals at $a_{exp}$. \label{CalcValsCoNi}}

    \begin{tabular}{l c c }
              & $m_{sp}$ ($\mu_B$/atom) & $E_{mag}$ (meV/atom) \\\toprule
     \multicolumn{3}{l}{hcp Co} \\
     PBE      &    1.65                & -255                 \\
     SCAN     &    1.80                & -578                 \\
     SCAN-L   &    1.63                & -277                 \\
              & & \\
          \multicolumn{3}{l}{fcc Ni} \\
     PBE      &    0.65                 & -60                  \\
     SCAN     &    0.78                 & -137                 \\
     SCAN-L   &    0.67                 & -74                  \\
              & & \\
          \multicolumn{3}{l}{bcc V} \\
     PBE      &    0.00                 &  0                  \\
     SCAN     &    0.57                 & -6                   \\
     SCAN-L   &    0.00                 &  0                  \\
\toprule     
    \end{tabular}
\vspace*{-12pt}
  \end{table}

For completeness, Table \ref{CalcValsCoNi} shows the calculated saturation
magnetization and FSM energies for hcp Co, fcc Ni and bcc V.
For Co, SCAN-L reduces
$m_{sp}$ relative to SCAN by about 9\%. For Ni the SCAN-L
reduction is 14\%.  The vanadium case is particularly notable,
because SCAN wrongly predicts a magnetic ground state, whereas
PBE and SCAN-L have non-magnetic ground states.

In summary, we have found that the SCAN over-magnetization 
arises from subtle differences in the
behavior of the iso-orbital indicator $\alpha$ for each
spin-channel that are magnified by the SCAN switching function, especially
for $\alpha < 1$. In contrast, the approximate nature of the 
orbital-independent $\alpha_L$ 
offsets those magnifications in a remarkably precise but serendipitous way.
Two approaches for the development of new meta-GGA functionals
are suggested by this analysis.  One would be to devise switching 
functions which are better adapted to the physics signaled by 
the iso-orbital indicator $\alpha$.  The other would be to
use a different iso-orbital indicator \cite{FurnessSun2019} 
that behaves more like
the approximate $\alpha_L$ in the cases considered.
One other possible consideration is that SCAN parameter values are determined, among
other constraints, by appropriate norms.  None of them is a
spin-polarized case.  As a consequence, SCAN relies solely on the 
spin-scaling relations for its magnetization predictions.  A new or
augmented set of norms might be useful.

\begin{acknowledgments}
We thank Fabien Tran for a helpful remark about WIEN2k 
and David Singh for informative discussion. This work 
was supported  by  U.S.\ Dept.\ of
Energy grant DE-SC 0002139.  SBT also was supported by 
U.S.\ Dept. of Energy Energy Frontier Research Center 
grant DE-SC 0019330.
\end{acknowledgments}

\end{document}